\documentclass{article}

\usepackage[preprint]{neurips_2019}

\usepackage{CJKutf8}
\usepackage[T1]{fontenc}    
\usepackage{hyperref}       
\usepackage{url}            
\usepackage{booktabs}       
\usepackage{amsfonts}       
\usepackage{nicefrac}       
\usepackage{microtype}      
\usepackage{graphicx}
\usepackage{epstopdf}
\usepackage{setspace}
\usepackage{rotating}
\usepackage{subfig}
\usepackage{verbatim} 
\usepackage{amssymb}
\usepackage{amsmath}
\usepackage{algorithm2e}
\usepackage{verbatim} 
\usepackage{epstopdf}
\usepackage{setspace}
\usepackage{amsthm}
\usepackage{rotating}
\usepackage[T1]{fontenc}
\usepackage{subfig}
\newtheorem{definition}{Definition}[section]

\usepackage{comment}
\usepackage{hyperref}
\usepackage[compact]{titlesec} 
\usepackage{CJKutf8}
\usepackage[T1]{fontenc}    
\usepackage{hyperref}       
\usepackage{url}            
\usepackage{booktabs}       
\usepackage{amsfonts}       
\usepackage{nicefrac}       
\usepackage{microtype}      
\usepackage{graphicx}
\usepackage{epstopdf}
\usepackage{setspace}
\usepackage{rotating}
\usepackage{subfig}
\usepackage{verbatim}
\usepackage{amssymb}
\usepackage{amsmath}
\usepackage{algorithm2e}
\usepackage{verbatim} 
\usepackage{epstopdf}
\usepackage{setspace}
\usepackage{amsthm}
\usepackage{rotating}
\usepackage[T1]{fontenc}
\usepackage{subfig}
\usepackage{comment}
\usepackage{hyperref}
\usepackage{tikz,xcolor,hyperref}
\usepackage{caption} 
\usepackage{hyperref}
\usepackage{setspace}
\usepackage{soul}
\usepackage{lineno}
\definecolor{lime}{HTML}{A6CE39}
\DeclareRobustCommand{\orcidicon}{
	\begin{tikzpicture}
	\draw[lime, fill=lime] (0,0) 
	circle [radius=0.16] 
	node[white] {{\fontfamily{qag}\selectfont \tiny ID}};
	\draw[white, fill=white] (-0.0625,0.095) 
	circle [radius=0.007];
	\end{tikzpicture}
	\hspace{-2mm}
}
\foreach \x in {A, ..., Z}{%
	\expandafter\xdef\csname orcid\x\endcsname{\noexpand\href{https://orcid.org/\csname orcidauthor\x\endcsname}{\noexpand\orcidicon}}
}

\title{Dinucleotide repeats in coronavirus SARS-CoV-2 genome: evolutionary implications}
\author{%
  Changchuan Yin  \orcidA{} \thanks{Correspondence author, cyin1@uic.edu}\\
     Department of Mathematics, Statistics, and Computer Science\\
   University of Illinois at Chicago\\
   Chicago, IL 60607\\
   USA\\
   \texttt{cyin1@uic.edu} \\
}
\begin{document}
\maketitle
\begin{abstract}
The ongoing global pandemic of infection disease COVID-19 caused by the 2019 novel coronavirus (SARS-COV-2, formerly 2019-nCoV) presents critical threats to public health and the economy since it was identified in China, December 2019. The genome of SARS-CoV-2 had been sequenced and structurally annotated, yet little is known of the intrinsic organization and evolution of the genome. To this end, we present a mathematical method for the genomic spectrum, a kind of barcode, of SARS-CoV-2 and common human coronaviruses. The genomic spectrum is constructed according to the periodic distributions of nucleotides, and therefore reflects the unique characteristics of the genome. The results demonstrate that coronavirus SARS-CoV-2 exhibits dinucleotide TT islands in the non-structural proteins 3, 4, 5, and 6. Further analysis of the dinucleotide regions suggests that the dinucleotide repeats are increased during evolution and may confer the evolutionary fitness of the virus. The special dinucleotide regions in the SARS-CoV-2 genome identified in this study may become diagnostic and pharmaceutical targets in monitoring and curing the COVID-19 disease.
\end{abstract}
\textbf{{\large keywords}}: COVID-19, SARS-CoV-2, genomic spectrum, dinucleotide repeats, evolution fitness, virulence
\section*{Highlights}
\begin{itemize}
	\item We present the genomic spectrum of the SARS-CoV-2 genome as a genomic signature. The genomic spectrum illustrates the periodic distributions of nucleotides in the genome. 
	\item SARS-CoV-2 genomic spectrum displays pronounced dinucleotide TT repeat islands in ORF1a (nsps 3-6).
    \item Dinucleotide TT repeat islands in ORF1a (nsps 3-6) of the SARS-CoV-2 genome correlate with evolution fitness and can be considered as the pathogen-associated molecular patterns (PAMPs).
\end{itemize}

\section{Introduction}
\label{Introduction}
The current global pandemic of COVID-19 caused by the novel coronavirus SARS-COV-2, formerly 2019-nCoV, has been a severe threat to public health and the economy since it emerged in Wuhan, China, December 2019. As of May 29, 2020, 5.7 million COVID-19 cases in the globe have been confirmed, and 357,688 deaths have occurred from COVID-19 disease \citep{who_2020}. SARS-CoV-2 is the aetiological agent and responsible for a large-scale outbreak of fatal disease. Understanding the notable features of the SARS-CoV-2 genome in zoonotic origin and the evolutionary trend is of importance for revealing the intervention targets, and disease control and prevention.

Coronaviruses (CoVs) are the largest group of enveloped, positive-sense, single-stranded RNA viruses. Taxonomically, coronavirus SARS-CoV-2 belongs to Nidovirales order, Coronaviridae family, Coronavirinae subfamily, and beta-CoV genus \citep{fehr2015coronaviruses}. SARS-CoV-2 is highly pathogenic with similar or lower pathogenicity as severe acute respiratory syndrome (SARS) coronavirus (SARS-CoV) in 2002–2003, and lower pathogenicity than Middle-East respiratory syndrome coronavirus (MERS-CoV) in 2012. Nevertheless, SARS-CoV-2 is highly transmissible in humans. To date, seven human CoVs (HCoVs) have been identified. Among them are alpha-CoVs, HCoV-229E, and HCoV-NL63. The remaining five beta-CoVs include HCoV-OC43, HCoV-HKU1, SARS-CoV, MERS-CoV, and SARS-CoV-2. Four common human CoVs, HCoV-229E, HCoV-OC43, HCoV-NL63, and HCoV-HKU1 usually cause mild symptoms, like the common cold and/or diarrhea \citep{su2016epidemiology}. In contrast, SARS-CoV/MERS-CoV, and current SARS-CoV-2 are particularly pathogenic, causing SARS and COVID-19, respectively.

Because of the resemblance of SARS-CoV-2 to SARS-like coronaviruses, bats are likely to act as natural reservoir hosts of the progenitor of SARS-CoV-2 \citep{andersen2020proximal}. It is established that SARS-CoV was transmitted from palm civet to humans \citep{wang2006review} and MERS-CoV from dromedary camels to humans, yet the zoonotic origin of SARS-CoV-2 is still unclear. The genome sequence analysis shows that SARS-CoV-2 has the closest relative SARS-like SLCoV/RaTG13, found in horseshoe bat (\textit{Rhinolophus affinis}) from Yunnan, China \citep{zhou2020pneumonia}. SARS-CoV-2 has 96.2\% overall genome sequence identity to SARS-like coronavirus SLCoV/RaTG13 \citep{zhou2020pneumonia}. Therefore, the natural reservoir of SARS-CoV-2 could be the horseshoe bat. Recent evidence suggests pangolins as host candidates \citep{lam2020identifying}, however, the host pangolins are not firmly determined. Not all bat SARS-like CoV can infect humans. For example, bat coronavirus, swine acute diarrhoea syndrome coronavirus (SADS-CoV) found in 2017 \citep{zhou2018fatal}, caused millions of piglet deaths, but no human cases. We may postulate that SARS-CoV-2 evolved from its progenitor through mutations and evolutionary fitness in the host-shift and adapting. SARS-CoV-2 are fast-evolving pathogens that continuously undertake mutations in the generations of infection of the host \citep{yin2020genotyping,wang2020decoding,korber2020spike}. This fact suggests that SARS-CoV-2 had evolved mutations in critical proteins and genome structures prior to establishing human infection. To understand how SARS-CoV-2 jumps from animals to adaptively infect humans is important to the surveillance of virus evolution and diversity, therefore ultimately controlling the COVID-19 pandemic and preventing future SARS-like outbreaks.

SARS-CoV-2 is an enveloped positive-strand RNA virus, having an exceptionally long (29.9kb) genome \citep{zhou2020pneumonia},. The genome consists of 5' leader cap sequence along with a 3' poly (A) tail, genes encoding non-structural proteins (nsps), and structural proteins, as well as several accessory proteins. Approximate two-thirds of the genome comprises two large overlapping open reading frames (ORF1a and ORF1ab), encoding polyproteins that are subsequently cleaved by viral proteases to generate 16 non-structural proteins (nsp1 to nsp16). Non-structural proteins are essential for RNA replication, transcriptions, and immune evasion. Nsp3, a large multi-domain and multi-functional protein, plays essential roles in virus replication. The papain-like protease (PLpro) activity of nsp3 is responsible for the initial processing of OFR1a protein. In addition, nsp3, together with nsp4 and nsp6, recruits intracellular membranes to form double-membrane vesicles (DVMs) to support viral RNA replication. Nsp5 is a second viral protease (3C-like protease, 3CLpro) that splits both ORF1a and ORF1ab proteins. The downstream regions of the genome encode structural proteins, the spike (S) protein, the nucleocapsid (N) protein, the envelope (E) protein, and the membrane (M) protein. The four structural proteins are all required to produce a structurally complete viral particle. The S protein mediates viral attachment to the ACE2 receptor host, and the subsequent fusion between the viral and host cell membranes enables the virus to enter host cells. The nucleocapsid (N) protein, one of the most abundant viral proteins, can bind to the RNA genome and participate in processes of replication, assembly, host cell response during viral infection \citep{mcbride2014coronavirus}.

The SARS-CoV-2 genome has been sequenced and annotated, but little is known for the complex structure of the genome in light of evolutionary fitness and host infectivity. Conventionally, RNA viruses use encoded proteins to interact with the components in cellular response. However, numerous discoveries show that the virus RNA structures, determined by genome composition may also play central roles in maximizing virus replication and evolutionary fitness \citep{jensen2012sensing}. For example, recent studies show that increased CG and TA dinucleotides in both coding and non-coding regions of echovirus 7 inhibit replication initiation during post-entry in several cell lines \citep{fros2017cpg}. Therefore, RNA viruses simulate host mRNA composition, for example, the dinucleotide frequencies \citep{fros2017cpg}. Animal genomes have a bias in their dinucleotide composition, and the heavy under-representation of CG and TA dinucleotides is especially well known. Most animal RNA and small DNA viruses suppress genomic CG and TA dinucleotide frequencies, apparently mimicking host mRNA composition \citep{di2017dinucleotide}. If a virus RNA composition or structure is very different from host mRNA, RIG-I (retinoic acid-inducible gene I)-like receptors may detect RNA molecules that are absent from the uninfected host \citep{goubau2013cytosolic}. Detecting evolutionary microbial structures known as pathogen-associated molecular patterns (PAMPs) is an important feature of the innate immune system. Host-cells possess intrinsic defense pathways that prevent replication of viruses with increased CG and TA frequencies in mechanisms independent of codon usage \citep{belalov2013causes}. 

The genomic spectrum demonstrates dinucleotide, trinucleotide, and multi-nucleotide distributions. The dinucleotide distributions are often considered as the signature of a genome \citep{kariin1995dinucleotide} and the 3-periodicity patterns are distinguishing characteristics for the protein-coding regions \citep{tsonis1991periodicity}. The strengths of the 2-periodicity and 3-periodicity in a genome are determined by the perfect levels and copy numbers of dinucleotide and trinucleotide repeats, respectively. Because 2-periodicity and 3-periodicity are the essential characteristics of a genome, in this study, we only examine these two periodicities from the genomic spectrum when analyzing the genome.

In this study, we present the genomic spectrum of SARS-CoV-2 and identify the dinucleotide repeats in ORF1a (nsps 3-6) that have been instrumental in interacting host immune systems and pathogenicity and host infectivity. These dinucleotide repeats are essential to the survival and infectivity of the microbe and can be considered as one of PAMPs in SARS-CoV-2. These genomic elements are evidence of the evolutionary fitness of SARS-CoV-2 in host-shift and adaption. Tracking the evolution of these elements may provide insights into the zoonotic origin of SARS-CoV-2 and the control of COVID-19 disease.

\section{Concepts and methods}
To inspect the insightful traits of the SARS-CoV-2 genome, we utilize our periodicity analysis method to survey the nucleotide distributions and the rendered periodicities in the genome. We previously proposed the periodicity analysis method to quantitatively detect the nucleotide repeats and periodicities in a genome \citep{yin2017identification}. The method employs nucleotide distributions on periodic positions in a genome and identifies approximate repeat structures as the signatures of the genome. Because we have included more functionalities, such as smoothing the periodicity profile, from the original method, here we describe the method in detail though the technical algorithms had been delineated previously \citep{yin2017identification}. Our computer programs of the periodicity analysis of a genome are available to the public at GitHub repository \href{https://github.com/cyinbox/DNADU}{https://github.com/cyinbox/DNADU}.

\subsection{Periodic nucleotide frequencies in a DNA sequence}
The nucleotide distributions at the periodic positions of a DNA sequence can be represented by a congruence derivative (CD) vector \citep{yin2016periodic,yin2017identification}. The CD vector of a nucleotide for a specific periodicity is constructed by the cumulative frequencies of the nucleotide at these periodic positions (Definition 2.1).

\begin{definition} For a DNA sequence of length $n$, let $u_\alpha  (k)=1$ when the nucleotide $\alpha$ appears at position $k$, otherwise, $u_\alpha (k)=0$, where $\alpha \in \{ A,T,C,G\}$ and $k = 1, \cdots , n$. The congruence derivative vector of the nucleotide $\alpha$ of the DNA sequence for periodicity $p$, is defined as\\
	\begin{equation}
	\begin{gathered}
	f_{\alpha ,j}  = \sum\limits_{\bmod (k,p) = j} {u_\alpha  (k)}  \hfill \\
	j = 1, \cdots ,p,k = 1, \cdots ,n \hfill \\ 
	\end{gathered} 
	\end{equation}
	, where $mod(k,p)$ is the modulo operation and returns the remainder after division of $k$ by $p$, and $f_\alpha  = \left( {f_\alpha (1),f_\alpha (2), \cdots ,f_\alpha (p)} \right)$.
\end{definition}

Four congruence derivative vectors $f_\alpha$ of periodicity $p$ for nucleotides A, T, C and G form a congruence derivative (CD) matrix of size $4 \times p$. The columns of the CD matrix indicate nucleotide frequencies at the periodic positions $k = pt-q$, where $k$ is the position index of a DNA sequence, $t = 1, 2, \ldots$, and $q = p - 1, \ldots , 2, 1, 0$. For example, consider the CD matrix of periodicity 5 for DNA sequence, the first column of the CD matrix shows the nucleotide frequencies at periodic positions $k = 1, 6, 11, \ldots , 5t - 4$; the second column of the matrix shows the nucleotide frequencies at periodic positions $k = 2, 7, 12, \ldots , 5t - 3$; the third column of the matrix shows the nucleotide distributions at periodic positions $k = 3, 8, 13, \ldots , 5t - 2$, and so on. The CD matrix of a DNA sequence describes nucleotide frequencies at all periodic positions and can be used to efficiently compute the Fourier power spectrum and determine periodicities in the DNA sequence \citep{yin2016periodic}. Therefore, the CD vector reflects the arrangement of repetitive sequence elements and inner periodicities in the DNA sequence. 

\subsection{The normalized distribution uniformity (NDU) of a DNA sequence}
Since the CD matrix contains the nucleotide frequencies on periodic positions, the variance of the matrix elements can measure the nucleotide distribution. For the CD matrix of periodicity $p$, the summation of $4p$ elements of the matrix is equal to the length $n$ of the DNA sequence, and the mean of the elements of the CD matrix is $\frac{n}{{4p}}$. Therefore, to quantify the nucleotide distribution, we define the normalized distribution uniformity (NDU) of a DNA sequence using the CD matrix (Definition 3.2).  

\begin{definition}
For a DNA sequence of length $n$, let $f_{i,j}$ be an element of the CD matrix of periodicity $p$, the normalized distribution uniformity of periodicity $p$ of the DNA sequence is defined as
\begin{equation}
NDU(p) = \frac{1}{n}\sum\limits_{i = 1}^4 {\sum\limits_{j = 1}^p {(f_{i,j}  - \frac{n}{{4p}})^2 } } 
	\end{equation}
\end{definition}
From Definition 2.2., we notice that the normalized distribution uniformity at periodicity $p$ is an intuitive description for the level of unbalance of nucleotide frequencies on periodic positions. It depends on the quadratic function of the nucleotide frequencies, sequence, and periodicity length. NDU(p) can be used to indicate the existence of the periodicity $p$ in a DNA sequence. This method offers an elaboration of the repetitive elements such as the repeat consensus, copy number, and the perfect level \citep{yin2017identification}. 

When using a sliding window along a genome, the periodicities of a range of 2 to 10 are calculated in each window segment. Therefore, a two-dimension periodicity spectrum is formed for the genome. The two-dimension spectrum can be considered as the genomics signature or the barcode signature of the genome. 

\subsection{Noise filtering and peaks detection}
To locate the positions of a repeat region in a genome, we smooth and filter the corresponding sliding-window periodicity by moving average convolution \citep{de1989smoothing}. Then the peaks in the periodicity profile are detected using the Z-score algorithm \citep{peaks2020}. The peak positions are used to demarcate repeats in a genome. 

In a nutshell, to compute distribution uniformities of different periodicities of a DNA sequence, we first scan the sequence in different periodicity sizes, construct the congruence derivative matrix of each periodicity, and compute the NDU(p) of these periodicities p. The periodicity with the maximum distribution uniformity reflects the predominant pattern of repetitive elements. The NDU values of periodicities indicate the perfect levels, and copy numbers of corresponding repeat regions. 

\subsection{Genome data}
This study depends on the complete genomes of coronaviruses, including SARS-CoV-2 \citep{wu2020new}, Severe Acute Respiratory Syndrome (SARS) related coronavirus, Middle East Respiratory Syndrome coronavirus (MERS-CoV), and human infection coronaviruses (human-CoVs). These genome data are retrieved from the National Center for Biotechnology Information (NCBI) Gene Bank. The bat SLCoV /RaTG13 complete sequence was reported by \citep{zhou2020pneumonia}, and downloaded from the GISAID repository (http://www.GISAID.org) \citep{shu2017gisaid}. The genome data in this study are listed in the supplementary material.

\section{Results and analysis}
\subsection{Genomic spectrum coronavirus SARS-CoV-2 reveals rich dinucleotide patterns in nsps}
To identify the signature features of the coronavirus SARS-CoV-2 genome, we employ the periodicity spectrum analysis to identify the characteristic periodicities in the genome. We create the genomic spectrum (barcode) of SARS-CoV-2 (Fig.1(a)) using the sliding window NDU method and compare it with the counterparts of SLCoV/RaTG13 (Fig.1(b)), SARS-CoV/Tor2 (Fig.1(c)), and MERS-CoV (Fig.1(d)).  From the spectrum comparison, we observe that SARS-CoV-2 and SLCoV/RaTG13 both have pronounced 2-periodicity in four regions while both SARS-CoV/Tor2 and MERS-CoV only have an extremely low level of 2-periodicity in the corresponding regions. The strong dinucleotide signal in SARS-CoV-2 encouraged us to investigate the causes in detail. 
\begin{figure}[tbp]
	\centering
	\subfloat[]{\includegraphics[width=2.75in]{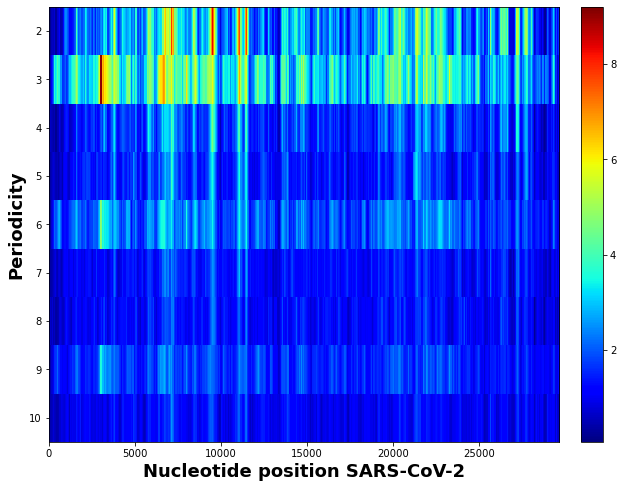}}
	\subfloat[]{\includegraphics[width=2.75in]{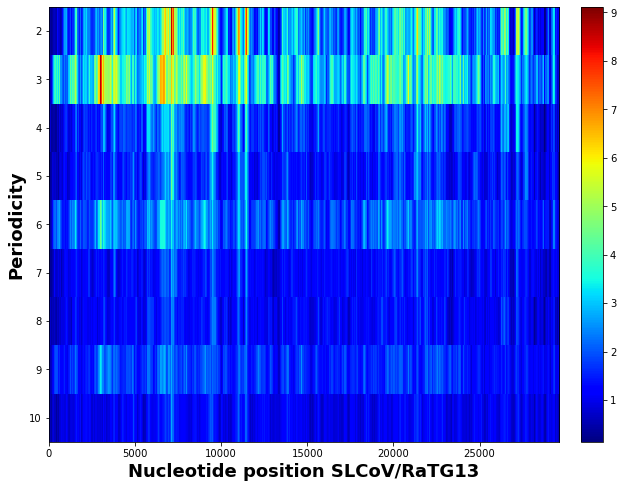}}\quad
	\subfloat[]{\includegraphics[width=2.75in]{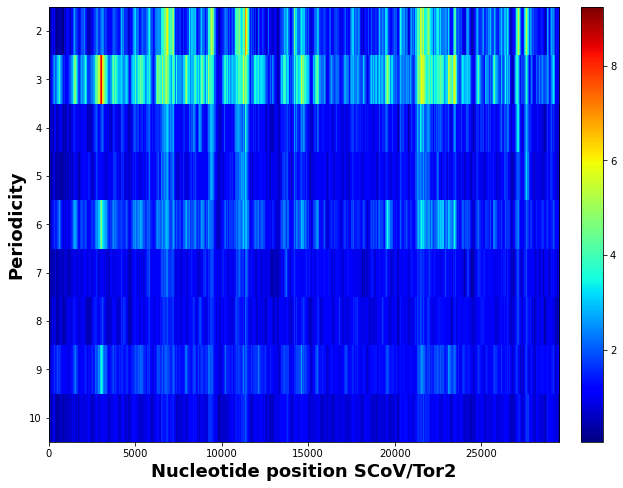}}
	\subfloat[]{\includegraphics[width=2.75in]{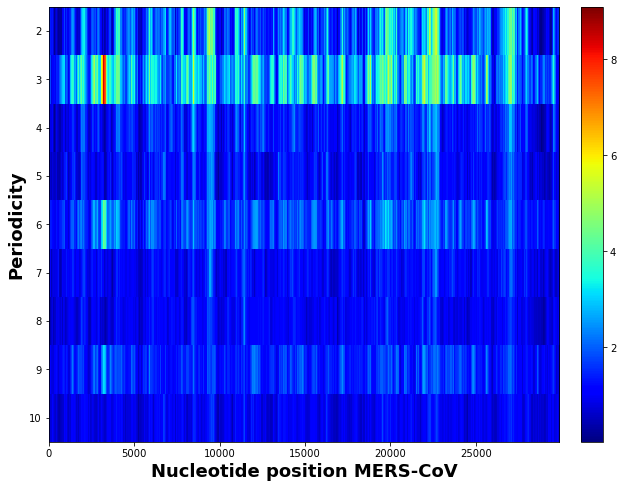}}\quad
	\caption{Genomic spectra of SARS-CoV-2 and SARS-related CoVs (SLoVs). (a) SARS-CoV-2. (b) SLCoV/RaTG13. (c) SARS-CoV/Tor2. (d) MERS-CoV. The sliding window is 250 bp.}
	\label{fig:sub1}
\end{figure}

To locate the regions of rich dinucleotide repeats, we verify that 2-periodicity and 3-periodicity are strong signals among all genomic periodicities (Fig. 2(a, d)), and detect the peaks of the sliding-window periodicity profiles (Fig.2 (b, c)). The peak positions are used to demarcate the dinucleotide repeat regions in the genome. The dinucleotide repeat regions (dinucleotide islands) are in ORF1a and the corresponding genes are listed in Table 1. 

\begin{figure}[tbp]
	\centering
	\subfloat[SARS-CoV-2/Wuhan-Hu-1]{\includegraphics[width=2.5in]{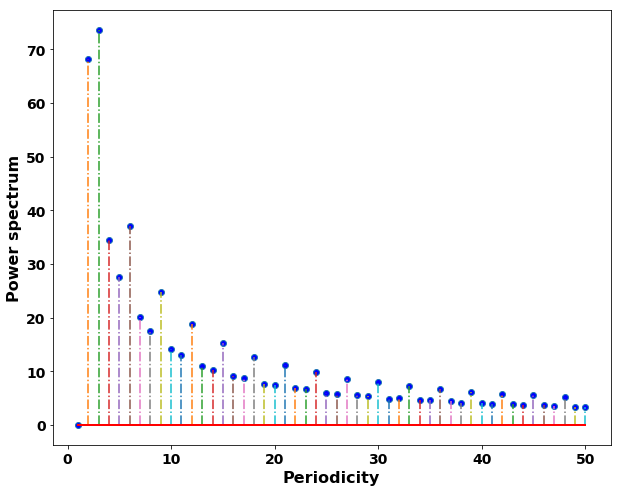}}
	\subfloat[SARS-CoV-2/Wuhan-Hu-1]{\includegraphics[width=2.5in]{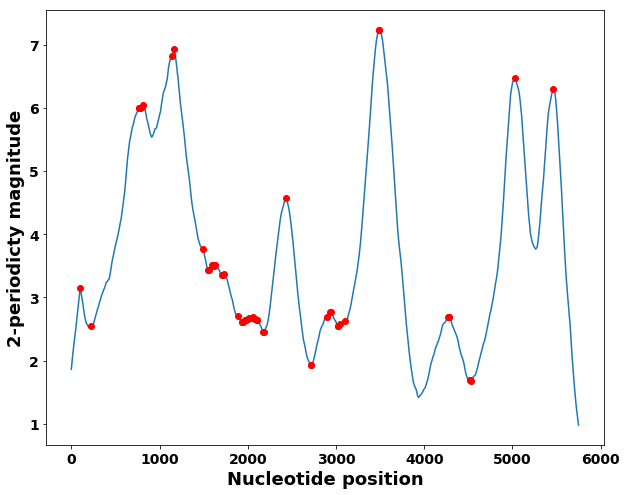}}\quad
	\subfloat[SARS-CoV-2/Wuhan-Hu-1]{\includegraphics[width=2.5in]{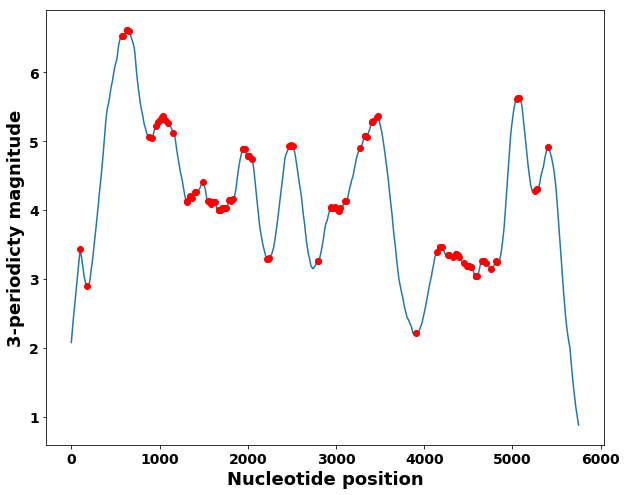}}
	\subfloat[SARS-CoV-2/Wuhan-Hu-1]{\includegraphics[width=2.5in]{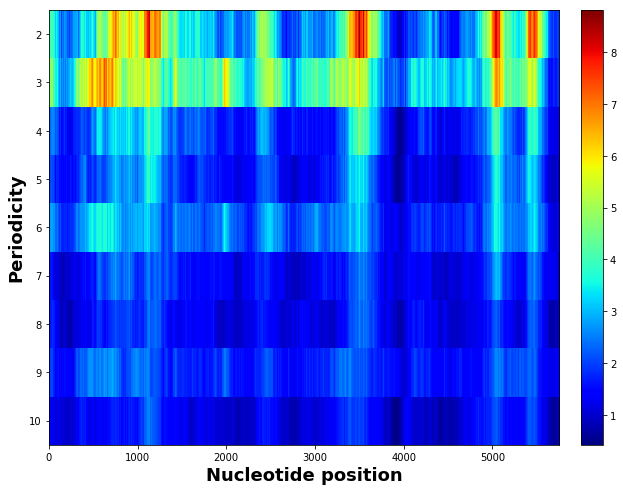}}\quad
	\caption{Genomic spectra of the region 6 kb - 12 kb of the SARS-CoV-2 genome (GenBank: NC\_0455127). (a) The periodicity magnitudes. (b) The 2-periodicity spectrum of 250 bp sliding windows. (c) The 3-periodicity spectrum of 250 bp sliding windows. (d) The genomic spectra of 250 bp sliding windows.}
	\label{fig:sub1}
\end{figure}

\begin{table}[ht]
	\caption{The dinucleotide repeats in SARS-CoV-2/Wuhan-Hu-1 genome (GenBank: NC\_045512)}
	\centering 
	\begin{tabular}{l*{4}{l}r}
		\hline\hline
		region & location & 2-periodicity &  consensus  & perfection level & protein \\
		\hline
		sub-region 1 &6227:7886 & 6.9320 & TT &  0.3599 & nsp3 \\
		sub-region 2 &9101:10273 & 7.2362 & TT & 0.3696 & nsp4 \\
		sub-region 3 &10850:12000 & 6.4728 & TT & 0.3783 & 3CLPro and nsp6 \\
		\hline\hline
	\end{tabular}
	\label{table:nonlin} 
\end{table}

From the genomic spectrum analysis, the relative abundance of the dinucleotide repeats, particularly dinucleotide TT, are mapped to the genes of ORF1a (nsp3, nsp4, and nsp6) (Fig.3 and Table 1.). However, these dinucleotide repeat signals are weak or imperceptible in the corresponding regions in the SARS-CoV and MERS-CoV genomes.

\begin{figure}[tbp]
	\centering
	{\includegraphics[width=4.75in]{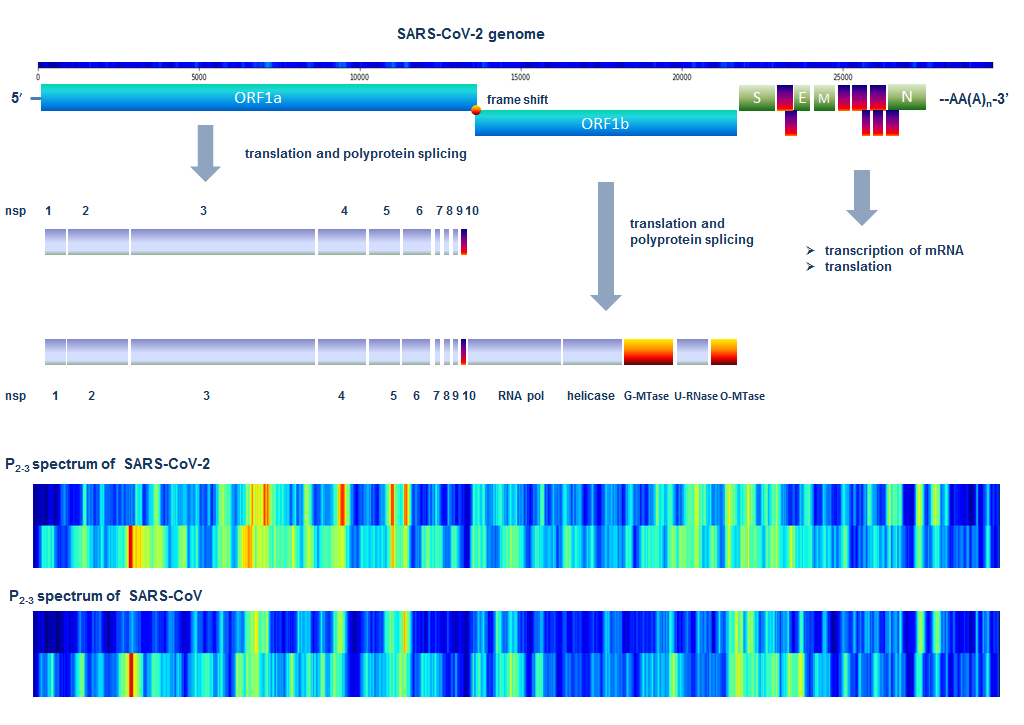}}
	\caption{Paradigm of dinucleotide repeats in the SARS-CoV-2 genome organization. The paradigm is drawn according to the reference genomes of SARS-CoV-2 (NC\_045512) and SARS-CoV/Tor2 (NC\_004718).}
	\label{fig:sub1}
\end{figure}

In the coronavirus SARS-CoV-2 RNA genome, the gene for replicase of 20 kb encodes two overlapping polyproteins, ORF1a (replicase 1a) and ORF1ab (replicase 1ab). The genome structure and the corresponding dinucleotide regions identified are illustrated in Fig.3. The two polyproteins are responsible for viral replication and transcription \citep{chen2020emerging}. The expression of the C-proximal portion of pp1ab requires (–1) ribosomal frame-shifting. The first dinucleotide repeat is in the coding region of Papain-like proteinase (PL proteinase, non-structural protein 3, nps3). Nsp3 is the largest essential component of the replication and transcription complex. The PL proteinase in nsp3 cleaves nsps 1-3 and blocks host innate immune response, promoting cytokine expression \citep{lei2018nsp3, serrano2009nuclear}. The second dinucleotide repeat is in the coding region of non-structural protein 4 (nsp4). Nsp4 is responsible for forming double-membrane vesicles (DMV). The third dinucleotide repeat is in the coding region of the C-terminal 3CLPro protease (3 chymotrypsin-like proteinase, 3CLpro) and nsp6. 3CLPro protease is essential for RNA replication. The 3CLPro proteinase is responsible for processing the C-terminus of nsp4 through nsp16 for all coronaviruses \citep{anand2003coronavirus}. Therefore, conserved structure and catalytic sites of 3CLpro may serve as attractive targets for antiviral drugs \citep{kim2012broad}. Together, nsp3, nsp4, and nsp6 can induce DMV \citep{angelini2013severe}.

In summary, the dinucleotide repeat islands found in this study are located in the host-interaction regions of the genome of SARS-CoV-2. These special dinucleotide repeat regions in ORF1a most likely contribute to the adaptive immune response, therefore, implying evolution fitness.  

Coincidentally, previous work on MERS-CoV using co-evolution analysis revealed that nsp3 represents a preferential selection target in adaptive evolution for zoonotic MERS-CoV to a new host \citep{forni2016extensive}. Our finding that nsp3 is involved in evolution fitness is consistent with the discovery in MERS-CoV. We investigate the correlation of dinucleotides AA and TA contents in SARS-CoV genomes and virulence. We examine the dinucleotide in the genomic regions (6 kb - 12 kb). The increased 2-periodicity in the genomic regions of SARS-CoV-2 and SARS-like CoVs are the results of the unbalanced distributions of dinucleotides.

\subsection{Dinucleotide repeats in major SARS-like coronaviruses (SLCoVs) in evolution}
To understand the evolutionary tendency of coronavirus genomes, we examine the genomic spectra of four major bats SARS-like coronaviruses (SLCoVs), pangolin-SLCoV, SLCoV/ZXC21, SLCoV/WIV1, and SLCoV/Shaanxi2011, all of which naturally live in bat \textit{Rhinolophidae horseshoe}. Because pangolin-SLCoV was found similar to SARS-CoV-2, pangolin was exploratorily postulated as an intermediate animal host of SARS-CoV-2 \citep{}. SLCoV/ZXC21 is the second similar strain to SARS-CoV-2, with 82\% similarity \citep{hu2018genomic}. SLCoV/WIV1 was closely related to SARS-CoV/Tor2 in terms of genome identity and ACE2 binding in human cells \citep{ge2013isolation}. SLCoV/Shaanxi2011 was found in 2011 \citep{yang2013novel}.

The results show that pangolin-SLCoV displays three major dinucleotide repeats in nsp3 and nsp4, but lacks the corresponding dinucleotide repeats in 3CLPro and nsp6 as found in SARS-CoV-2 (Fig.4(a)). So SARS-CoV-2 is mostly closed to SLCoV/RaTG13 (Fig.1(b)) and SLCoV/ZXC21 (Fig.4(b)), not pangolin-SLCoV. If pangolins are the intermediary hosts of SARS-CoV-2, and SARS-CoV-2 was indeed evolved from pangolin-SLCoV, we may infer that pangolin-SLCoV would need to evolve dinucleotide repeats in 3CLPro and nsp6 during evolution fitness before infecting human hosts. 

We also observed the evolution trend between SLCoV/WIV1 (Fig.4(c)) and SLCoV/Shaanxi2011 (Fig.4(d)). The genomic spectrum of SLCoV/Shaanxi2011 is similar to SLCoV/WIV1, but has additionally increased dinucleotide repeat in 3CLPro and nsp6. This new dinucleotide repeat in the region 3CLPro and nsp6 in SLCoV/Shaanxi2011 is consistent with the regions found in SARS-CoV-2, SLCoV/RaTG13, and SLCoV/ZXC21. Therefore, the dinucleotide repeat in the region 3CLPro and nsp6 probably play an important role in the evolutionary fitness of SARS-CoV-2 to the human hosts.  

The results show that only SARS-CoV and SARS-like CoVs have low dinucleotide repeats. The low-dinucleotide contents can be considered as in early evolution fitness when interacting with the human immune system, then low-dinucleotides may render high virus virulence because the virus has not adapted to the host immune system, and the host immune system acts intensely. 

\begin{figure}[tbp]
	\centering
	\subfloat[]{\includegraphics[width=2.75in]{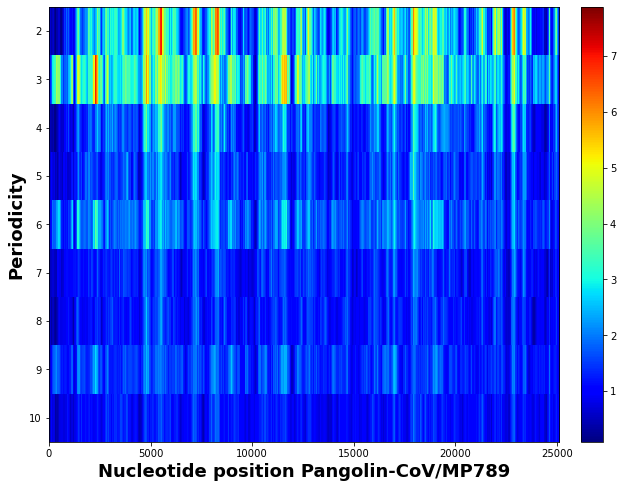}}
	\subfloat[]{\includegraphics[width=2.75in]{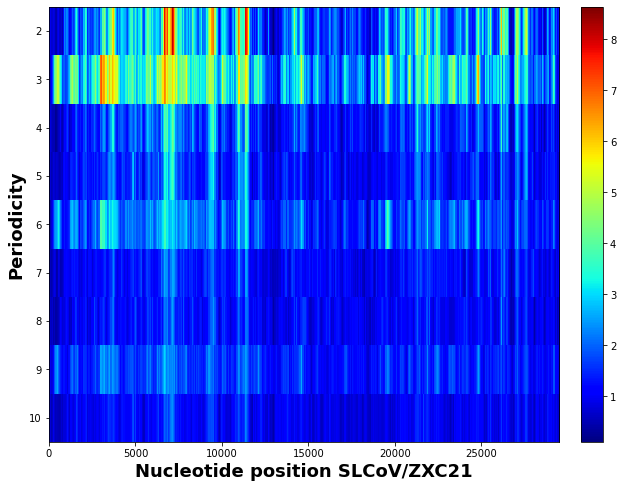}}\quad
	\subfloat[]{\includegraphics[width=2.75in]{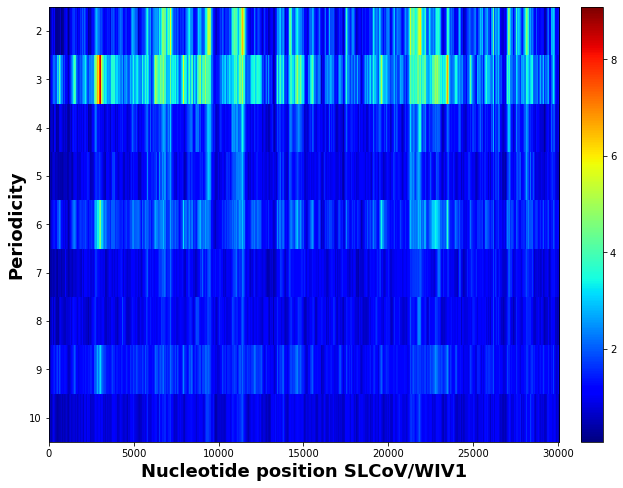}}
	\subfloat[]{\includegraphics[width=2.75in]{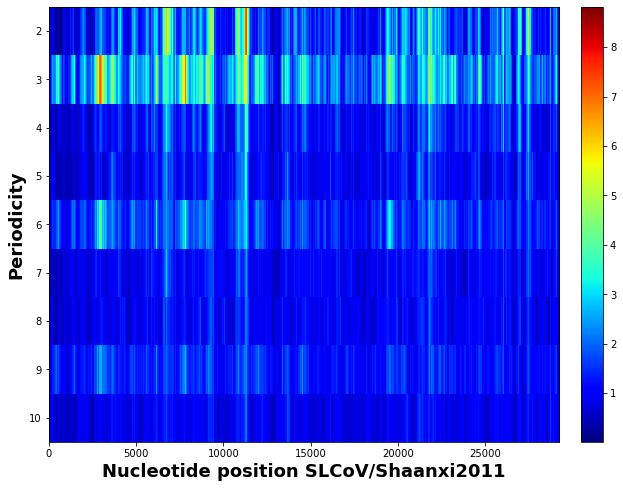}}\quad
	\caption{Genomic spectra of SARS-like CoVs (SLCoVs). (a) pangolin-SLCoV. (b) SLCoV/ZXC21. (c) SLCoV/WIV1. (d) SLCoV/Shaanxi2011. The sliding window is 250 bp.}
	\label{fig:sub1}
\end{figure}

\subsection{Trend of dinucleotide repeats in human coronaviruses during evolution}
To investigate the correlation of dinucleotide repeats and pathogenicity of coronaviruses, we produce and compare the genomic spectra of four common human coronaviruses (Fig.5). Classical human coronavirus 229E (HCoV-229E) and human coronavirus OC43 (HCoV-OC43) were identified in 2004. The two viruses are close relatives, and the virus characteristics are similar to human pathogenicity. Both HCoV-229E and HCoV-OC43 can cause young children and the elderly and have a low immune function. Almost 100\% of children are infected in early childhood, mainly as self-limiting upper respiratory infections, such as the common cold and intestinal infections Symptoms caused by HCoV-OC43 strain are generally more severe than those of HCoV-229E virus. From the genomic spectrum analysis, we observe higher 2-periodicity in HCoV-229E than in HCoV-OC43 (Fig.5 (a,c)). These dinucleotide repeats regions correspond to the three sub-regions in SARS-CoV-2. High 2-periodicity value may attenuate the virus replication and therefore reduce severe virulence. That is in an agreement with the correlation of dinucleotide repeats and pathogenicity previously. 

The spectra of HCoV-NL63 and HCoV-HKU1 demonstrate extremely high 2-periodicity, as well as 3-periodicity in the corresponding regions (Fig.5 (b,d)). HCoV-NL63 and HCoV-HKU1 are the most common human CoVs that cause only a mild cold symptom or no symptom \citep{pyrc2007novel}. Again, these high dinucleotide repeats may contribute to the light pathogenicity in these two viruses.  

\begin{figure}[tbp]
	\centering
	\subfloat[]{\includegraphics[width=2.75in]{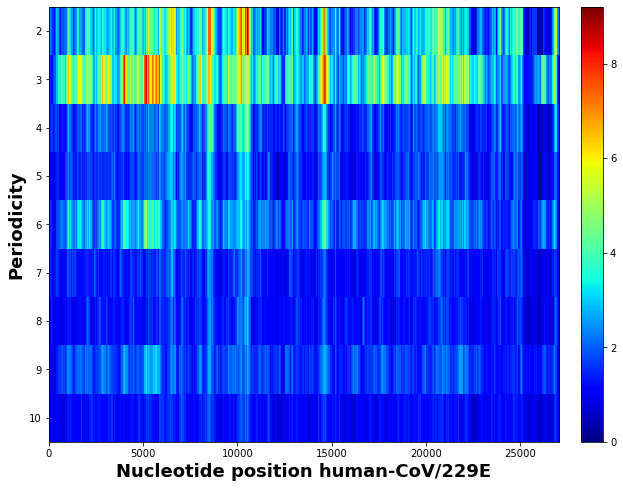}}
	\subfloat[]{\includegraphics[width=2.75in]{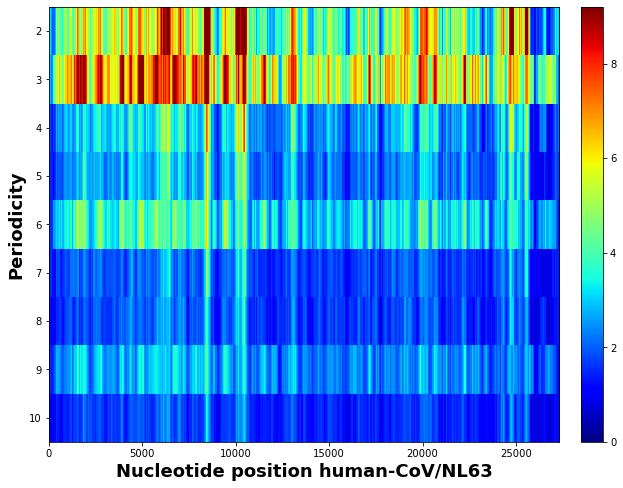}}\quad
	\subfloat[]{\includegraphics[width=2.75in]{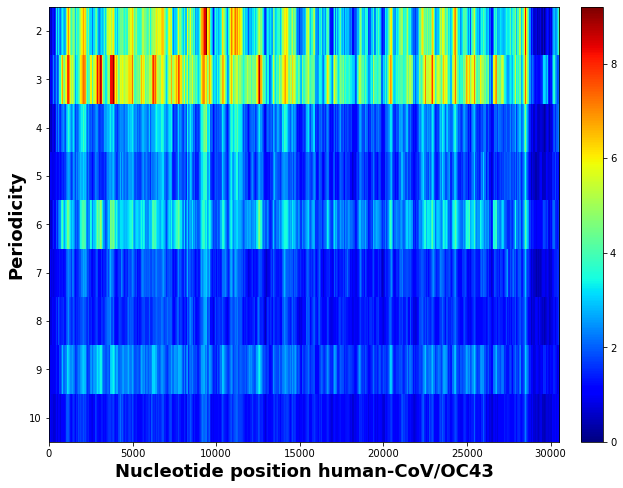}}
	\subfloat[]{\includegraphics[width=2.75in]{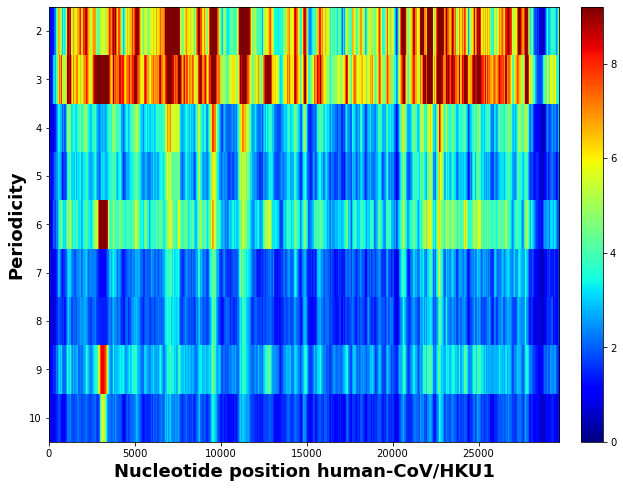}}\quad
	\caption{Genomic spectra of common human coronaviruses. (a) HCoV-229E. (b) HCoV-NL63. (c) HCoV-OC43. (d) HCoV-HKU1. The sliding window is 250 bp.}
	\label{fig:sub1}
\end{figure}

\subsection{Dinucleotide frequencies of SARS-CoV-2 during evolution}
We wish to know whether SARS-CoV-2 originally evolved from SARS-CoV, or SARS-CoV-2 would evolve to SARS-CoV. The answer to this question may help us to predict the evolution of SARS-CoV-2 virus for better disease prediction and control. Because the genomes of SARS-related coronaviruses over a long time period are rarely available, to infer the evolution of coronaviruses, we track the trend of HCoV-229E coronaviruses over the last six decades from the first human infected HCoV-229E identified in 1962 \citep{thiel2001infectious}. HCoV-229E virus causes common cold but occasionally it can be associated with more severe respiratory infections in children, elderly, and persons with underlying illness. Using the measurements of dinucleotides in the genomic regions, the trend of HCoV-229 coronaviruses may infer the evolutionary stages of bat coronaviruses. In a similar method, we may then determine the origin SARS-CoV-2 if the trend of SARS-CoV-2 is compared with SARS-like coronaviruses.

The human coronavirus HCoV-229E strains used in the dinucleotide trend analysis are from different historical periods. The reference genome HCoV-229 in the evolutionary analysis was obtained from the infectious HCoV-229E, the 1973-deposited laboratory-adapted prototype strain of HCoV-229E (VR-740). The HCoV-229E prototype strain was originally isolated in 1962 from a patient in Chicago. The first clinical HCoV-229E isolate from a US patient in 2012 was included in this study \citep{farsani2012first} (GenBank: JX503060). HCoV-229E (SC3112) isolate in 2015 is included (GenBank: KY983587). The new Human coronavirus strain HCoV-229E was isolated from plasma collected from a Haitian child in 2016 \citep{bonny2017complete} (GenBank: MF542265).

The result in the periodicity trends of the HCoV-229E coronaviruses demonstrates that both 2-periodicity (Fig.7(a)) and 3-periodicity (Fig.7(b)) in the coronaviruses are increasing with evolutionary time. SARS-CoV-2 has relatively high 2-periodicity and 3-periodicity. This result suggests that the trends of 2-periodicity and 3-periodicity are increasing with time. The evolutionary origin of coronaviruses can be inferred by the trends of 2-periodicity and 3-periodicity. Therefore, we may compare the 2-periodicity and 3-periodicity in SARS-CoV-2 and SARS-like coronaviruses to understand the evolutionary origin of SARS-CoV-2.
\begin{figure}[tbp]
	\centering
	\subfloat[2-periodicity]{\includegraphics[width=4.75in]{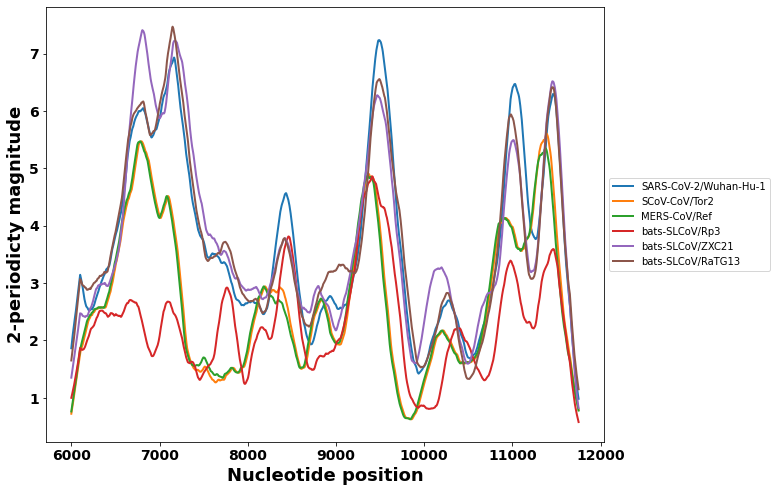}}\quad
	\subfloat[3-periodicity]{\includegraphics[width=4.75in]{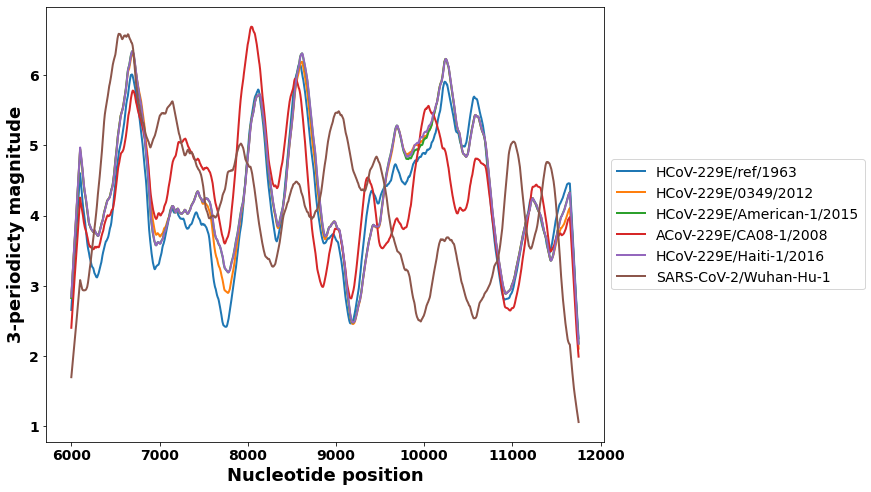}}\quad
	\caption{Comparison of periodicities in genomic regions (6k-12k) of human HCoV-229E coronaviruses. (a) 2-periodicity. (b) 3-periodicity.}
	\label{fig:sub1}
\end{figure}
To investigate the correlation of dinucleotide repeats with virus virulence in SARS-related coronaviruses, we compare the spectra of the genomic region at coordinates 6k-12k bp of five SARS-related coronaviruses. The genomic regions contain abundant dinucleotide repeats. The region 6k-12k of the genome contains three dinucleotide repeats approximately located at 6k-8k, 8k-10k, and 10k-12k sub-regions. The five SARS-related coronaviruses have different levels of virulence. The most virulent virus is MERS-CoV, followed by SARS-CoV. The 2-periodicity magnitudes, which reflect the distribution of dinucleotides, are compared and shown in (Fig.8 (a,b)). We may see that MERS-CoV genomic region has the lowest dinucleotide level in all three dinucleotide sub-regions. SARS-CoV also has a low dinucleotide level but is higher than MERS-CoV. We, therefore, may infer that the low dinucleotide level correlates with high virulence. The lower the dinucleotide level is, the higher virus virulence is. This postulation is supported by the observation of the dinucleotide in three SARS-related coronaviruses. Compared with the SARS-CoV, the SARS-like bats-SLOV/Rp3 has similar dinucleotide distributions in sub-regions 1 (6k-8k) and 3 (10k-12k), but higher dinucleotide distribution in sub-region 2 (8k-10k). SLCoV/Rp3 has lower virus virulence than SARS. The SARS-like SLCoV/ZXC21, which shares the highest sequence identity with SARS-CoV-2, shows slightly lower dinucleotide distributions in the two sub-regions 2 and 3, and slightly higher dinucleotide distribution than sub-region 1. 

We notice that trends of the 2-periodicity and 3-periodicity from MERS-CoV, SARS-CoV, SARS-like CoVs, and SARS-CoV-2 are increasing (Fig.8 (a,b)). Based on the previous analysis of the periodicity trends of coronaviruses at different times, we may infer that SARS-CoV-2 originates from the SARS-like CoVs.

\begin{figure}[tbp]
	\centering
	\subfloat[2-periodicity]{\includegraphics[width=4.75in]{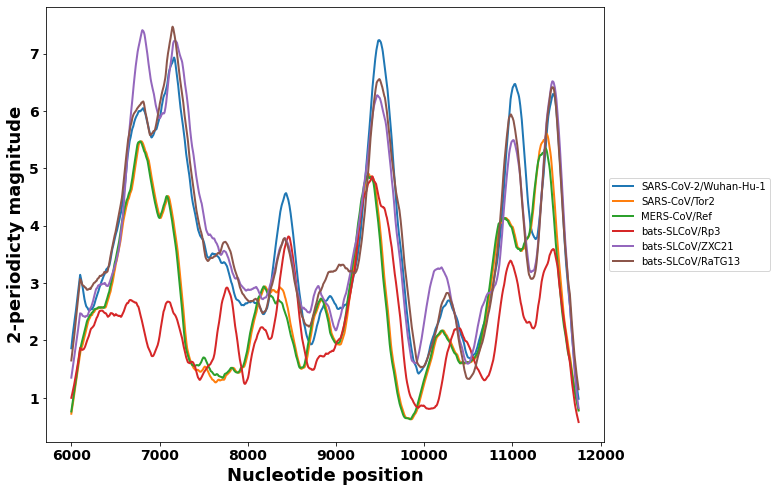}}\quad
	\subfloat[3-periodicity]{\includegraphics[width=4.75in]{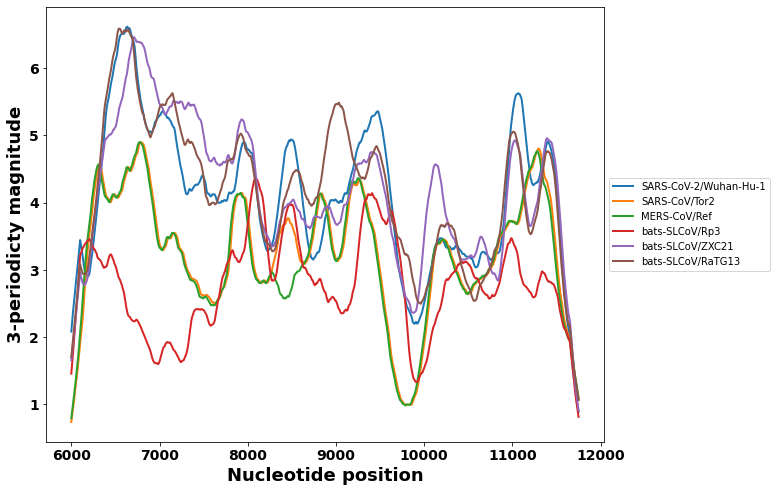}}\quad
	\caption{Comparison of periodicities in the genomic regions (6 kb- 12kb) of SARS-CoV and SARS-like coronaviruses. (a) 2-periodicity. (b) 3-periodicity.}
	\label{fig:sub1}
\end{figure}

\subsection{The dynamics of dinucleotide repeats in SARS-CoV-2 during evolution}
It is well studied that dinucleotide composition bias in RNA viruses may impact the virus replications, specifically, attenuating or strengthening virus virulence during evolution \citep{fros2017cpg,gu2019dinucleotide}. Clinical evidences have suggested that SARS-CoV-2 has lower virulence than SARS-CoV. To investigate if the increased dinucleotide repeats correlate with virus virulence, we compare the dinucleotide frequencies in the three dinucleotide rich islands of SAR-CoV-2 and SARS-CoV/Tor2. 

The result shows that both SARS-CoV-2 and SARS-CoV have an abundance of dinucleotides TT and TA in the whole genome, and three prominent dinucleotide repeat regions (Fig.9 (a,b,c,d)), and extreme CG deficiency in sub-region 3 (Fig.9(d)). The dinucleotides TT and TA are increased in SARS-CoV-2 compared with SARS-CoV (Fig.9 (a,b,c,d)).  

\begin{figure}[tbp]
	\centering
	\subfloat[whole genome]{\includegraphics[width=2.75in]{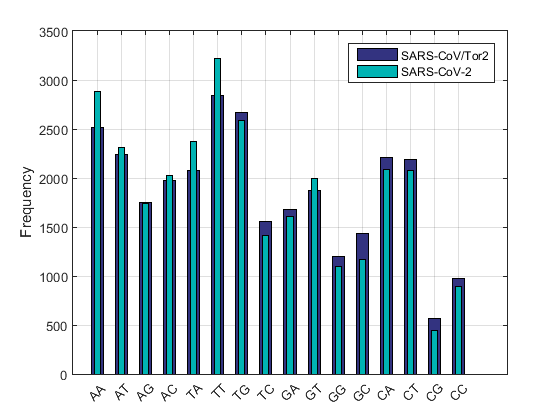}}
	\subfloat[sub-region 1]{\includegraphics[width=2.75in]{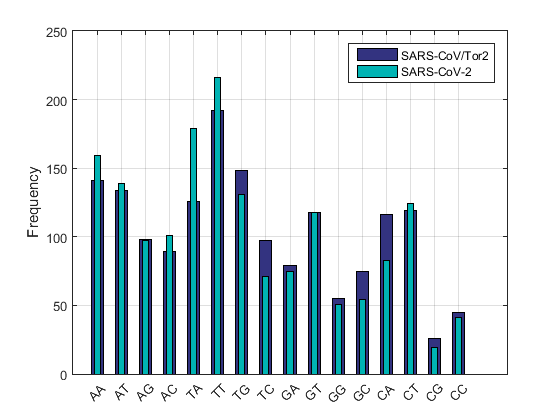}}\quad
	\subfloat[sub-region 2]{\includegraphics[width=2.75in]{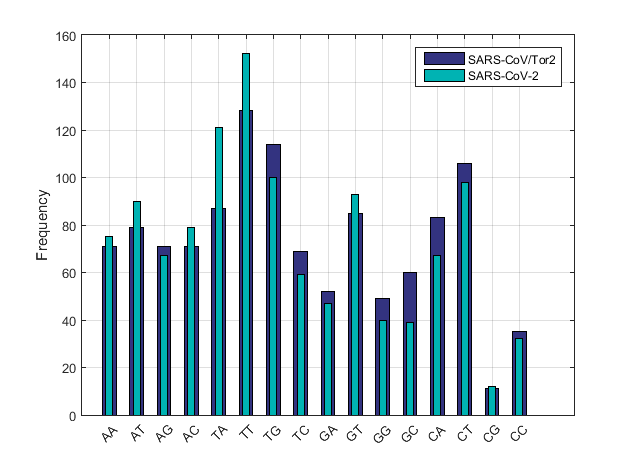}}
	\subfloat[sub-region 3]{\includegraphics[width=2.75in]{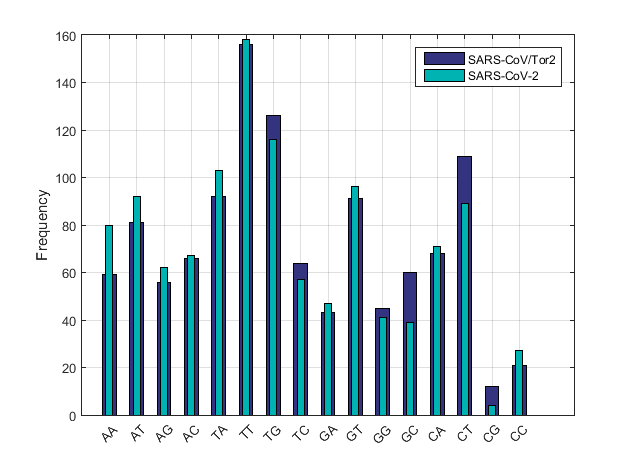}}\quad
	\caption{The dinucleotide frequency distributions in the three genomic regions of SARS-CoV-2 and SARS-CoV/Tor2. (a) whole genome, (b) sub-region 1. (c) sub-region 2. (d) sub-region 3. The sub-regions are listed in Table 1.}
	\label{fig:sub1}
\end{figure}

The role increased dinucleotides TT and TA in the SARS-CoV-2 genome is to possibly attenuate virus replication. One mechanism for attenuating virus by dinucleotide bias is that the dinucleotide regions fold a special structure as a target for cell RNA cleavage, which is a fundamental host response for controlling viral infections \citep{zhou1993expression}. The 2,5-oligoadenylate synthetase/RNase L system is an innate immunity pathway that responds to a pathogen-associated molecular pattern (PAMP) to induce degradation of viral and cellular RNAs, thereby blocking viral infection. In higher vertebrates, this process is often regulated by interferons (IFNs). Ribonuclease L (RNase L, L is for latent) is an interferon (IFN)-induced antiviral ribonuclease which, upon activation, destroys all RNA within the cellular and viral \citep{silverman2007viral}. RNase L cleaves hepatitis C virus (HCV) RNA at single-stranded TT and TA dinucleotides throughout the open reading frame (ORF). An interesting discovery is that in bacterium \textit{Mycoplasma pneumonia}, which is a respiratory infection agent, the genomes also have relative abundance extremes dinucleotides TT and TA \citep{karlin1998global}. Therefore, we may postulate that the dinucleotides TT and TA regions in SARS-CoV-2 are possibly cleaved by RNase L during infection.

\section{Discussion and perspectives}
In this study, we identify the unique dinucleotide repeats in the SARS-CoV-2 genome. The dinucleotide repeats in the genomic spectra are revealed by the periodicity analysis. We discover the strength of these repeats correlates with the evolutionary fitness of the virus to a human host for maximizing its survival in epidemics, instead of destroying the host. Therefore, RNA viruses simulate host mRNA composition such as the dinucleotide compositions \citep{fros2017cpg}. Most vertebrate RNA and small DNA viruses suppress genomic CG and TA dinucleotide frequencies, apparently mimicking host mRNA composition. The abundance of dinucleotides TT and TA are most likely common pathogenicity islands in microbial genomes. This study on SARS-CoV-2 provides additional evidence that increased dinucleotides TT and TA in SARS-CoV-2 is the result of interaction with the host during virus evolution. We consider these three dinucleotide abundance regions as pathogenicity islands of the SARS-CoV-2 genome. In addition, these special regions may contribute RNA replications, and can be recognized by cell RNAase L for RNA degradation in the immune response. However, this study is only a theoretical analysis of the genomes. The actual functional consequences and the impacts on transmissibility and pathogenesis of these dinucleotide repeats should be determined by biochemical experiments and animal models.

In humans and mammals, APOBEC (apolipoprotein B mRNA editing enzyme, catalytic polypeptide-like3) systems help protect the organisms from viral infections. For the origin of dinucleotide repeats in the SARS-CoV-2 genome, we speculate that the molecular mechanism of increased dinucleotide repeats during evolution fitness is possibly APOBEC3-mediated editing of viral RNA, in which Cytosine (C) is often mutated to Uracil (U) by deamination\citep{bishop2004apobec}. Therefore, many dinucleotide TT repeats could be generated as the 2-periodicity island in the RNA genome by the APOBEC defense system. 

Cross-species transmission of coronaviruses from wildlife reservoirs may lead to disease outbreaks in humans, posing a severe threat to human health. To date, most studies on the zoonotic origin of SARS-CoV-2 primarily focus on the Spike protein, which is essential for the entry of virus particles into the cell. Mutations or acquisition of potential cleavage site for furin proteases in the Spike protein may confer zoonotic transmissibility of SARS-CoV-2 \citep{andersen2020proximal,hoffmann2020multibasic}. However, the spike protein is required but not sufficient for zoonotic coronavirus transmission. For instance, the Spike protein in SARS-like SHC014-CoV can only enable the chimeric virus SHC014-MA15 to infect human cells when the Spike protein is integrated into a wild-type SARS-CoV backbone \citep{menachery2015sars}. The SARS-CoV backbone contains ORF1ab replication components. Our study provides evidence for the importance of ORF1ab replication regions in evolutionary fitness. Consequently, these ORF1ab replication components are critical for zoonotic transmissions. 

This study on the genomic spectrum of SARS-CoV-2 reveals high dinucleotides TT regions in ORF1a might contribute to evolution fitness in host immune evasion. Accordingly, monitoring SARS-CoV-2 and developing antiviral drugs should envisage molecular characteristics and changes of nsps 3-6, which are vital components in interacting with human hosts. These special dinucleotide repeat regions should be investigated in detail for its functions and phenotype changes in SARS-CoV-2. Importantly, these dinucleotide repeat regions can possibly be the prophylactic and therapeutic targets for controlling COVID-19.

\section*{Acknowledgments}
The author sincerely thanks the researchers worldwide who sequenced and shared the complete genomes of SARS-CoV-2 and other coronaviruses from GISAID (https://www.gisaid.org/). This research is dependent on these precious data. The author specially appreciates Prof. Jiasong Wang (Nanjing University, China), Dr. Gang Cheng (University of Illinois at Chicago), and Dr. Guo-Wei Wei (Michigan State University) for valuable suggestions.

\section*{Abbreviations}
\begin{itemize}
	\item ACE2: angiotensin converting enzyme 2 
	\item APOBEC: apolipoprotein B mRNA editing enzyme, catalytic polypeptide-like3 
	\item COVID-19: coronavirus disease 2019 
	\item CD: a congruence derivative
    \item DMV: double-membrane vesicle	
    \item MERS: Middle-East respiratory syndrome
    \item NDU: the normalized distribution uniformity
	\item PAMPs: pathogen-associated molecular patterns
	\item RIG-I: retinoic acid-inducible gene I
    \item SARS: severe acute respiratory syndrome
	\item SARS-CoV-2: severe acute respiratory syndrome coronavirus 2
\end{itemize}
\section*{Supplementary materials}
\begin{table}[ht]
	\caption{The genomes of coronaviruses used in this study.}
	\centering 	
	\begin{tabular}{l*{4}{l}r}
		\hline\hline
		GenBank/GISAID & Name \\
		\hline
		NC\_045512   & SARS-CoV-2/Wuhan-Hu-1    \\ 
		AY274119 & human-SCoV/Tor2    \\ 
		NC\_019843 & MERS-SCoV/Ref           \\ 
		AY278489 & human-SCoV/GD01     \\ 
		AY297028 & human-SCoV/ZJ01    \\  
		DQ071615 & bats-SLCoV/Rp3    \\
		MN996532 & bats-SLCoV/RaTG13    \\  
		EPI\_ISL\_410542& Pangolin-CoV/Guangxi/P2V/2017\\
		MT084071 & pangolin-CoV/MP789\\
		MG772933 & bats-SLCoV/ZC45   \\ 
		MG772934 & bats-SLCoV/ZXC21    \\ 
		JX993987 & bats-SLCoV/Rp-Shaanxi2011    \\ 
		KC881005 & bats-SLCoV/RsSHC014   \\   
		KC881006 & bats-SLCoV/Rs3367   \\     
		DQ412043 & bats-SLCoV/Rm1   \\ 
		DQ648856 & bats-SLCoV/273-2005   \\ 
		DQ648857 & bats-SLCoV/279-2005   \\ 
		DQ084200 & bats-SLCoV/HKU2-2    \\ 
		DQ084199 & bats-SLCoV/HKU2-3   \\ 
		KF367457 & bats-SCoV/W1V1    \\ 
		AY613947 & civets-SCoV/GZ0402  \\ 
		AY304486 & civets-SCoV/SZ3  \\ 
		AY613950 & civets-SCoV/PC4    \\ 
		NC\_002645 & human-CoV/229E   \\ 
		NC\_005831 & human-CoV/NL63   \\ 
		AY585229 & human-CoV/OC43   \\ 
		AY597011 & human-CoV/HKU1  \\ 
		MH940245 & human-CoV/HKU1\_2  \\ 
		NC\_009988 & bats-CoV/HKU2     \\ 
		MG557844 & swine-SADS-CoV    \\ 
		NC\_003045 & bovine-Cov    \\
		AF304460 & HCoV-229E/ref/1963  \\ 
		JX503060 & HCoV-229E/0349/2012  \\ 
		JX503061 & HCoV-229E/J0304/2012  \\ 
		KY983587 & HCoV-229E/American-1/2015 \\ 
		KY684760 & HCoV-229E/American-2/2015  \\ 
		KY967357 & HCoV-229E/American-3/2015 \\ 
		MF542265 & HCoV-229E/Haiti-1/2016  \\ 
		\hline\hline
	\end{tabular}
	\label{table:nonlin} 
\end{table}

\clearpage
\bibliographystyle{elsarticle-harv}
\bibliography{../References/myRefs}
\end{document}